\documentstyle[sprocl,epsfig]{article}

\def\ap#1#2#3   {{\em Ann. Phys. (NY)} {\bf#1} (#2) #3}
\def\apj#1#2#3  {{\em Astrophys. J.} {\bf#1} (#2) #3}
\def\apjl#1#2#3 {{\em Astrophys. J. Lett.} {\bf#1} (#2) #3}
\def\app#1#2#3  {{\em Acta. Phys. Pol.} {\bf#1} (#2) #3}
\def\ar#1#2#3   {{\em Ann. Rev. Nucl. Part. Sci.} {\bf#1} (#2) #3}
\def\cpc#1#2#3  {{\em Computer Phys. Comm.} {\bf#1} (#2) #3}
\def\err#1#2#3  {{\it Erratum} {\bf#1} (#2) #3}
\def\ib#1#2#3   {{\it ibid.} {\bf#1} (#2) #3}
\def\jmp#1#2#3  {{\em J. Math. Phys.} {\bf#1} (#2) #3}
\def\ijmp#1#2#3 {{\em Int. J. Mod. Phys.} {\bf#1} (#2) #3}
\def\jetp#1#2#3 {{\em JETP Lett.} {\bf#1} (#2) #3}
\def\jpg#1#2#3  {{\em J. Phys. G.} {\bf#1} (#2) #3}
\def\mpl#1#2#3  {{\em Mod. Phys. Lett.} {\bf#1} (#2) #3}
\def\nat#1#2#3  {{\em Nature (London)} {\bf#1} (#2) #3}
\def\nc#1#2#3   {{\em Nuovo Cim.} {\bf#1} (#2) #3}
\def\nim#1#2#3  {{\em Nucl. Instr. Meth.} {\bf#1} (#2) #3}
\def\np#1#2#3   {{\em Nucl. Phys.} {\bf#1} (#2) #3}
\def\pcps#1#2#3 {{\em Proc. Cam. Phil. Soc.} {\bf#1} (#2) #3}
\def\pl#1#2#3   {{\em Phys. Lett.} {\bf#1} (#2) #3}
\def\prep#1#2#3 {{\em Phys. Rep.} {\bf#1} (#2) #3}
\def\prev#1#2#3 {{\em Phys. Rev.} {\bf#1} (#2) #3}
\def\prl#1#2#3  {{\em Phys. Rev. Lett.} {\bf#1} (#2) #3}
\def\prs#1#2#3  {{\em Proc. Roy. Soc.} {\bf#1} (#2) #3}
\def\ptp#1#2#3  {{\em Prog. Th. Phys.} {\bf#1} (#2) #3}
\def\ps#1#2#3   {{\em Physica Scripta} {\bf#1} (#2) #3}
\def\rmp#1#2#3  {{\em Rev. Mod. Phys.} {\bf#1} (#2) #3}
\def\rpp#1#2#3  {{\em Rep. Prog. Phys.} {\bf#1} (#2) #3}
\def\sjnp#1#2#3 {{\em Sov. J. Nucl. Phys.} {\bf#1} (#2) #3}
\def\zp#1#2#3   {{\em Z. Phys.} {\bf#1} (#2) #3}

\newcommand{\ad}{{\gamma_{0}^{2}}}
\newcommand{\xt}{X_{\perp}}

\begin{document}

\title{MULTIPARTICLE PRODUCTION IN THE SOFT LIMIT AND 
QCD COHERENCE\footnote{Talk (based on paper ref. [6]) 
presented at the 8th Int. Workshop on Multiparticle Production, 
''Correlations and
Fluctuations '98 '', June 14-22, 1998, Matrahaza,
Hungary, to be publ. in the Proc. by World Scientific, Singapore.}}

\author{WOLFGANG OCHS}

\address{Max-Planck-Institut f\"ur Physik, F\"ohringer Ring 6,\\
D-80805 Munich, Germany\\e-mail: wwo@mppmu.mpg.de}

\maketitle\abstracts{ The production of gluons in a jet 
is considered in limited phase space, either with a cut 
in transverse momentum 
with respect to the jet axis $k_\perp<k_\perp^{cut}$
or with a cut in absolute momentum
$|\vec{k} | <k^{cut}$. It is shown in the perturbative QCD calculations
that in the soft limit $k_\perp^{cut} \to 0$ the multiplicity distribution
becomes Poissonian, whereas for $k^{cut} \to 0$ the distribution remains
non-Poissonian. The Poissonian limit is a consequence of the soft gluon
coherence in the genuine multiparticle correlations.
We also investigate how incoherent
hadronization processes could possibly modify the predictions for small cut-off
parameters.}

\section{Introduction}
Multiparticle production remains an interesting testing ground for
the predictions of perturbative QCD and also for the models of 
colour confinement. An approach of great simplicity is based on the idea of
duality between the properties of the hadronic and partonic final state.
This became first quantitative for the particle energy spectra which -- when
calculated in the modified leading log approximation (MLLA) --  
provided a good description of the data \cite{adkt} (``Local Parton Hadron
Duality''). This poses the question whether the perturbative calculations
are relevant to a larger class of observables including the soft particle
production. Quite a number of different observables have been calculated%
\cite{dkmt} and the comparison with experiment is encouraging \cite{ko}.

The standard hadronization models and the duality picture are comparable in
the first phase of the jet evolution where secondary partons are produced by
bremsstrahlung and pair production processes down to a relative scale of
about 1 GeV. At this scale nonperturbative processes may take over as they
are described in hadronization models. Alternatively, the 
perturbative phase may be evolved 
further down towards lower scales of a few 100 MeV
before the cascade is terminated by a cut-off in transverse momentum
\begin{equation}
k_\perp \geq Q_0.  \label{cutoff}
\end{equation}
 This non-perturbative cut-off represents the influence of the confinement:
only if a gluon is radiated with larger transverse momentum it gives rise to
the production of an extra particle. The production of hadrons may then be
rather similar to the production of partons and in fact the data on hadron
and jet multiplicities suggest  taking the number of partons
and hadrons to be equal \cite{lo}.

In order to test this duality picture further it is particularly important
to investigate the final stage of the jet evolution at scales around and
even below 1 GeV. Can the decay processes involving resonance decays be
described in an average sense by a perturbative partonic cascade? 

To approach this question it is interesting to study the peculiar aspects of
perturbative QCD in the soft region, in particular:\\
1. the running of the coupling $\alpha_s(k_\perp)$ with the transverse
momentum\\
2. the soft gluon coherence which limits the emission angle of a soft gluon
by the angle of the nearest colour connected parton.\\
Some support for the running coupling at small scales is suggested by the
rapid rise of multiplicity under small scale variations at small absolute
scales~\cite{lo}. Support for the coherence phenomenon at small scales
can be drawn from the observation of an essentially energy independent 
yield of soft particles (below a momentum of $\sim$200 MeV) over the full
energy range explored in $e^+e^-$ annihilation \cite{klo}. 

In this talk I will report about a recent analysis 
of genuine multiparticle correlations in the soft limit 
carried out in collaboration with S. Lupia and J. Wosiek \cite{low}.
The question is
whether the peculiar predictions of the perturbative theory are
followed by hadrons in this novel situation.
As a consequence of the soft gluon coherence \cite{ao} the  gluons 
emitted with small transverse momentum 
from the primary parton in the jet -- which typically requires small angles --
 cannot radiate again. Therefore in this limit the gluon emission resembles
the photon emission in QED in Born approximation and one expects besides the
energy independent bremsstrahlung with a flat rapidity plateau an
independent multiparticle emission and this corresponds to a Poissonian
distribution of the emitted soft particles.

\section{Definition of Observables and Methods of calculation}
In order to study the multiparticle phenomena in the soft limit we introduce
the multiplicity moments in restricted phase space
\begin{equation}   
f^{(q)}_c(C,Q)=\int_{\Gamma_c(C,Q)} \rho^{(q)}(\vec{k}_1\dots \vec{k}_q;Q)
d \vec{k}_1\dots d \vec{k}_q , \label{cut}
\end{equation}
where $\rho^{(q)}$ denotes the $q$-particle inclusive distribution 
in the momenta $k_i$ at energy scale $Q$ and
the phase space integration is restricted by $\Gamma_c(C,Q)$
with a cut variable $C$. 
Two types of cuts 
to be applied to all particles in the
final state have been discussed\cite{low}: 
the momentum cut $|\vec
{k}_i| < k^{cut}$,
and the transverse momentum cut $k_{\perp,i} < k_{\perp}^{cut}$,  
for $i=1,...,q$.
Clearly, the moments (\ref{cut}) determine the multiplicity distribution
of particles produced in the restricted phase space, and as such  they
provide a more differential characteristics than the global quantities.
The latter ones are retained for 
the maximal $k^{cut}$ and $k_\perp^{cut}$ at a given  $Q$.

The theoretical analysis in perturbative QCD 
has been carried out in two ways. First, analytical results have been
obtained within the double logarithmic approximation (DLA) \cite{bcm,dfk}
in which only the contributions of the leading soft and
collinear singularities are kept. These results can  serve to provide the 
correct qualitative picture and to test the influence of the colour
coherence and the running coupling on the results. The DLA for the $cut$
moments has been worked out completely and the results can be obtained
perturbatively to any order in the coupling~\cite{low}.

It is known since a while
that the global multiplicity moments are not reproduced by the 
DLA results quantitatively. More recently it has been shown\cite{OPAL} 
that these moments are not well reproduced by the 
NLO results \cite{mw} neither nor  by the NNLO calculations \cite{dln} 
although a clear improvement is then obtained. Agreement with the data
is found finally in a numerical solution of the evolution equations \cite{sl}.  
In order to obtain realistic estimates of the new $cut$ moments we have
therefore derived also the predictions from a Monte Carlo program (ARIADNE
\cite{ariadne}) which is built on similar procedures than
 the analytical calculations. 

\section{Analytical results on Multiplicity Moments}\label{sec:small}

\subsection{Results in the soft limit}
The theoretical analysis starts from the equations for the generating
functionals for the quark and gluon jets; then by the appropriate
differentiation one obtains the equations for cumulant and factorial moments
which are solved and the moments integrated over the regions of phase space
required.

In the case of small cut-off parameters which is 
of  primary interest to us various simplifications in the analytical 
analysis can be applied: first, it is enough to take into account the terms
of lowest order in the coupling, moreover, in this limit, the solutions for
running coupling approach those for fixed coupling in which case close 
 expressions are obtained. We present here only the main results and refer
to the original paper \cite{low} for the derivation.

Let's consider first the cylindrical phase space with small cut-off
$k_\perp^{cut}$. 
For the normalized cumulant moments we find
\begin{equation}
K^{(q)}(X_\perp,Y) =
\frac{qf_{q-1}}{2q-1} \left(\frac{X_\perp}{Y}\right)^{q-1}
\quad {\rm for}\quad X_\perp\ll\lambda,\ Y. \label{Kqsx}
\end{equation}
in terms of the logarithmic variables
\begin{equation}
X_\perp = \ln \frac{k_\perp^{cut}}{Q_0},\quad Y=\ln\frac{P\Theta}{Q_0},\quad
\lambda=\ln\frac{Q_0}{\Lambda},\quad  \label{logvar}
\end{equation}
for a jet of momentum $P$ and half opening angle $\Theta$.
The numbers $f_q$ are to be calculated recursively $(f_0=1,\
f_1=\frac{1}{2},\ f_2=\frac{1}{3},\ldots)$.
One observes that the moments $K^{(q)}$ decrease quickly with the order $q$ 
 corresponding to what has been called ``linked pair ansatz''
\cite{ces} for multiparticle correlations. For the normalized factorial
moments ($F^{(2)}=1+K^{(2)}=<n(n-1)>/<n>$)
one obtains for small cut-off $X_\perp$
\begin{equation}
F^{(q)}(X_\perp,Y) \cong 1 + \frac{q(q-1)}{6} \frac{X_\perp}{Y}.
       \label{FqLO}
\end{equation}

So we obtain the remarkable result that for small transverse momentum
cut-off all factorial moments approach unity and therefore the multiplicity
distribution becomes Poissonian. This is a consequence of the dominance of
the single soft gluon emission at small $k_\perp$, i.e., the absence of
branching processes with secondary gluon emissions. This behaviour is
just analogous to the usual QED bremsstrahlung and follows from the
coherence of the soft gluon radiation and the angular ordering
condition \cite{ao} which limits the angles of the secondary particle emission
by the (typically small) emission angle of the first gluon.
Namely, if this angular
ordering condition is suppressed artificially the moments remain constant
with $X_\perp$ and the multiplicity distribution remains non-Poissonian.

On the other hand, if we carry out the same calculations for the spherically
cut moments we obtain quite a different behaviour: the moments stay
 constant down to small cut-off parameters
$X=\ln\frac{k^{cut}}{Q_0}$ as
\begin{equation} 
K^{(q)}(X,Y) = 2^{q-1} f_{q-1}/(2q-1). 
\label{Kk}      
\end{equation}      
Consequently, soft gluons with limited
momentum $k$ 
have essentially a non-Poissonian multiplicity distribution, while   
those with limited $k_\perp$ follow a Poissonian one. 
These calculations can be carried out for running $\alpha_s$ as well and the
results approach those of fixed $\alpha_s$ in the soft limit as expected on
general grounds.

\subsection{General Solution in the DLA}
 The multiplicity moments to arbitrary order in the coupling $\alpha_s$ 
can be obtained by solving evolution equations
in the full range of the cut-off's $0<X,X_\perp<Y$.
The results reduce to the solutions for small cut-off discussed above and
approach correctly the known results for the global moments in full phase
space.

The evolution equations for the unnormalized cumulant  moments
$c^{(q)}=K^{(q)} \bar n^q$ in case of
spherical momentum cut read
\begin{eqnarray}
c^{(q)}_{sph}(X,Y)& = & 
\int_0^X dy \ad(y)[qf^{(q-1)}(y)+f^{(q)}(y)](X-y) 
\nonumber \\&+&
\int_0^X dx\int_x^{Y-X+x} dy \ad(y)f^{(q)}_{sph}(x,y)
%
%
\label{cqmom}
\end{eqnarray}
  where $f^{(q)}(Y)$ refer to the known global unnormalized factorial moments
and $\gamma_0^2=6\alpha_s/\pi$.
From this equation the $cut$ moments
can be calculated recursively to any precision.

The moments in cylindrical  phase space can be obtained from those in
spherical phase space using the equation 
\begin{eqnarray}
\lefteqn{ 
         c^{(q)}_{cyl}(\xt,Y)= c^{(q)}(\xt)+} \nonumber \\&&
(Y-\xt)\int_0^{\xt} \ad(y)
[qf^{(q-1)}(y)+f^{(q)}(y)]dy+ \nonumber \\&&
\int_{\xt}^Y (Y-y) \ad(y) f^{(q)}_{sph}(\xt,y) dy . \label{cqrel}
\end{eqnarray}

\begin{figure}[t]
\begin{minipage}{.95\linewidth}
 \begin{center}
\mbox{\epsfig{file=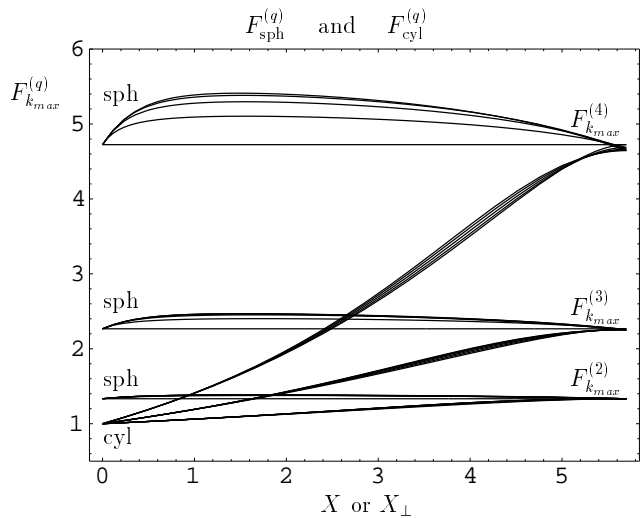,bbllx=0.cm,bblly=5cm,bburx=21.cm,bbury=21cm,width=10cm}}
          \end{center}
\end{minipage}
\caption{ 
DLA predictions for the cut-off dependence of the first
 three normalized moments $F^{(q)}_{k_{max}},\;q=2,3,4$ for $Y=5.7$.
 Both families, i.e. spherically (sph) and cyllindrically (cyl) cut moments,
 are shown.              
 They coincide in the global limit (X or $X_{\perp} = Y$), but have
distinctly
different threshold behaviour. Different lines which describe one moment
correspond to different orders $k_{max}$ of the perturbation theory included
$k_{max}=q,q+1,...,8$.
} 
\label{figtheory} 
\end{figure}

The results from this calculation up to $O(\alpha_s^8)$ 
are shown in Fig.~\ref{figtheory} 
 in the full range of cut-off parameters $X,X_\perp$. The figure displays
the very different behaviour of the moments in spherical and cylindrical
phase space with the latter ones approaching unity for $X_\perp\to 0$
corresponding to the Poisson limit.

\section{Monte Carlo results}
\begin{figure}[ht]
\vspace{-1.6cm} 
 \begin{minipage}{.95\linewidth}
          \begin{center}
\mbox{\mbox{\epsfig{file=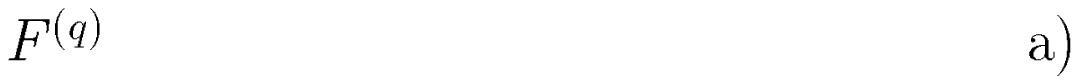,bbllx=5.cm,bblly=20.cm,bburx=5.2cm,bbury=24.cm,width=0.15cm}}
\mbox{\epsfig{file=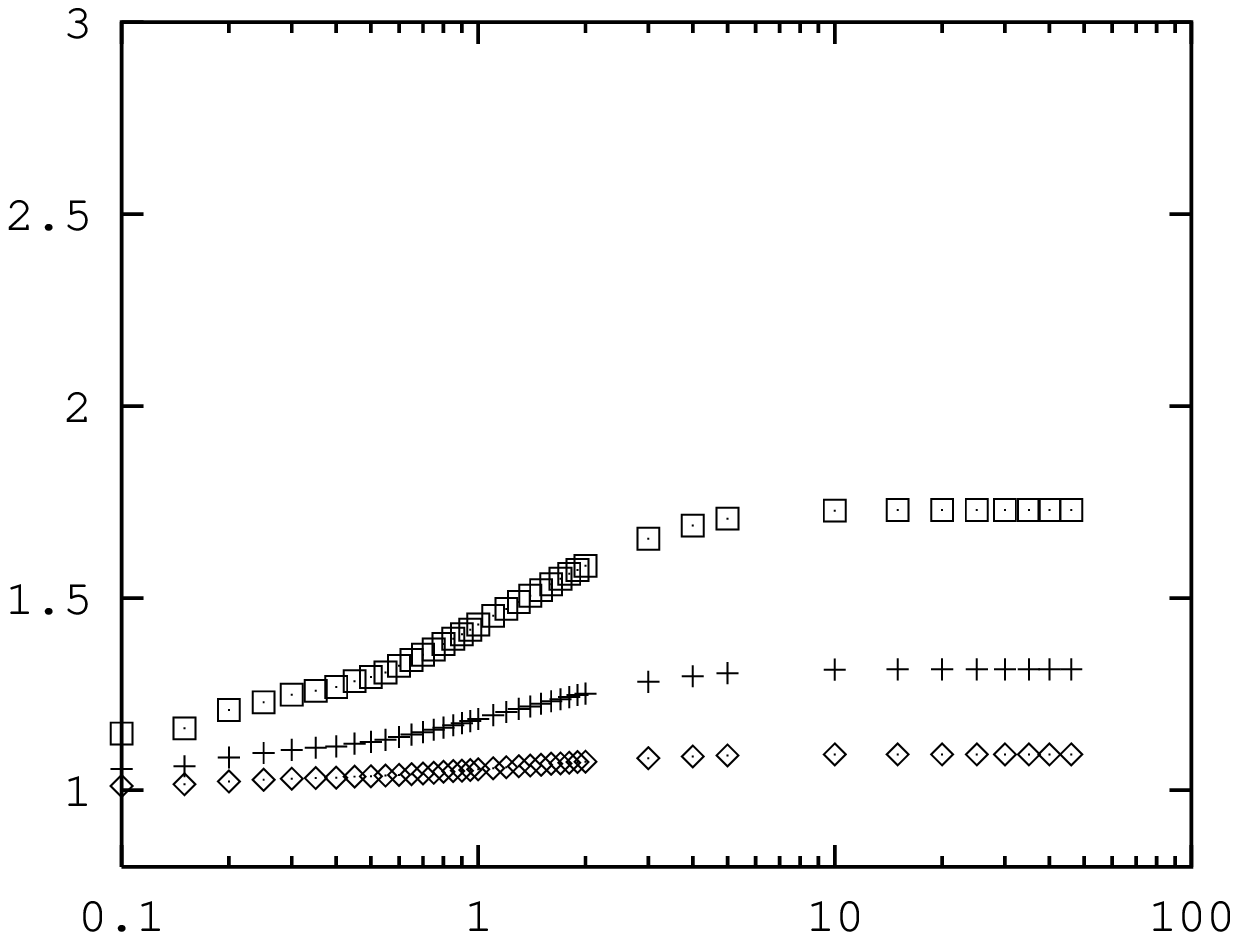,bbllx=3.cm,bblly=2.5cm,bburx=17.cm,bbury=14.5cm,width=9.cm}}
      }         \end{center}
\vspace{-0.2cm}
\hspace{5.5cm}  $k_\perp^{\mathrm cut}$ [GeV]
\vspace{-0.4cm}
      \end{minipage}
        \vspace{0.5cm}
 \begin{minipage}{.95\linewidth}
          \begin{center}
\mbox{\mbox{\epsfig{file=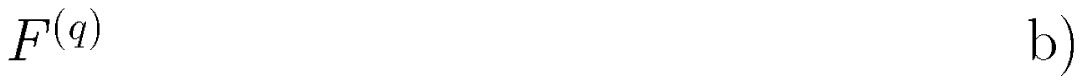,bbllx=5.cm,bblly=20.cm,bburx=5.2cm,bbury=24.cm,width=0.15cm}}
\mbox{\epsfig{file=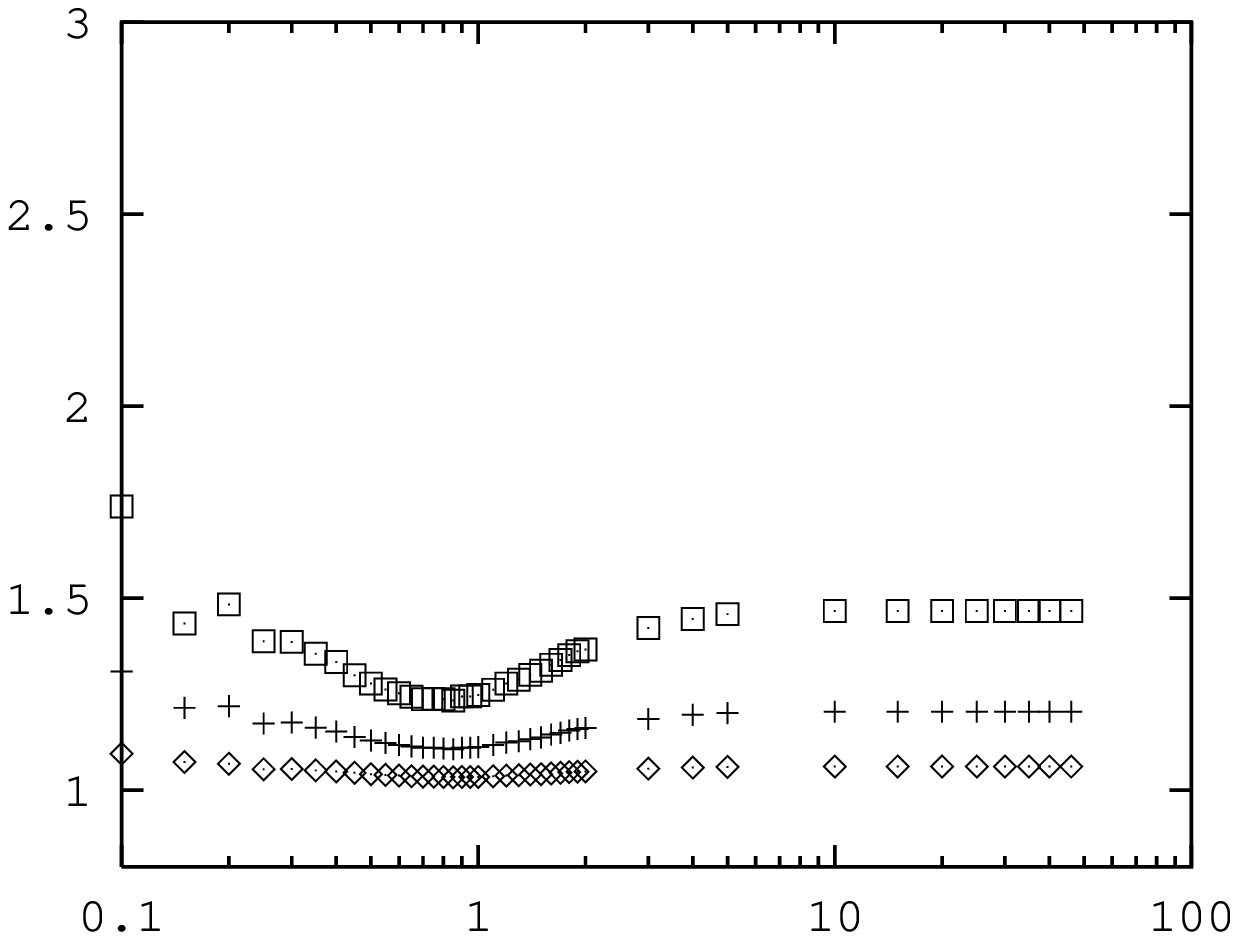,bbllx=3.cm,bblly=2.5cm,bburx=17.cm,bbury=14.5cm,width=9cm}}
      }           \end{center}
\vspace{-0.2cm}
\hspace{5.5cm}  $k_\perp^{\mathrm cut}$ [GeV]
      \end{minipage}
\caption{(a) cut moments of order 2 (diamonds), 3 %
 (crosses) and 4 (squares) in one hemisphere defined through the thrust
axis as a function of $k_\perp^{cut}$ as predicted
~by ARIADNE at parton level with parameters readjusted; 
(b) same as in (a), but at	
hadron level with default
values of the parameters and string fragmentation.}              
\label{figmathra}
\end{figure}

We complement our analytical DLA calculations by a Monte Carlo
computation  of the multiplicity moments.
Our aim is to clarify whether the qualitative features of our DLA
calculations are verified and to obtain quantitative predictions
at the parton level.
We have chosen the ARIADNE program \cite{ariadne} which has a similar
procedure in evolving the parton cascade and in the final cutoff
(\ref{cutoff}) which can be chosen independently of the QCD scale $\Lambda$.
By comparing the Monte Carlo results with the data on the total
multiplicities and global moments in $e^+e^-$ annihilation we adjust the
parameters of the program  $Q_0=0.2$ GeV and $\lambda=0.015$.
With these parameters we obtain the parton level predictions for the
cylindrical moments in Fig. \ref{figmathra}a. The moments clearly show the
decrease towards unity for decreasing cut parameter $k_\perp^{cut}$ in
qualitative 
agreement with our analytical calculation. The absolute size, however, is lower in the
Monte Carlo than in the analytical calculation which neglect various finite
energy effects. The moments for the spherical cut (not shown) keep some
larger non-zero values at small cut-off. We take these findings as confirmation
of our main analytical results on the very different behaviour of the two
families of moments and of the approach to the Poissonian limit of the
cylindrical moments.

Furthermore, we studied the effect of hadronization on the considered
observables. The idea of Local Parton Hadron Duality \cite{adkt} 
has been originally
proposed for the single inclusive distributions. The study of the moments
\clearpage
\noindent
considered here allows for a novel test of these ideas in the case of genuine
multiparticle correlations. In Fig. \ref{figmathra}b we show the results of
the ARIADNE program with string hadronization using the standard parameters.
There is the remarkable prediction that the moments first 
decrease with decreasing
$k_\perp^{cut}$ as in the parton level calculation but for
 $k_\perp^{cut}<0.8$  GeV  the moments rise again and do not approach the
Poissonian limit. 


\section{Conclusions and Outlook to Other Processes}

We have shown that perturbative QCD predicts the different behaviour of
multiplicity moments in cylindrical and spherical phase space regions. The
approach to a Poissonian in the soft limit in case of the cylindrical cut is
a consequence of the soft gluon coherence. This behaviour is also supported
by the Monte Carlo calculation which takes into account the most important
non-leading effects. The Poissonian multiplicity corresponds to the
independent emission of soft gluons distributed with a flat rapidity plateau
in close similarity to the QED Bremsstrahlung.
On the other hand, hadronization effects may distort the
predicted behaviour. It will therefore be very interesting to study these
moments experimentally and to find out whether, or, 
 to what extent the perturbative
predictions survive the hadronization process in this case of genuine
multiparticle correlations.

Our calculations apply to quark and gluon jets in hard collisions.
If the (untriggered) high energy $pp$ collisions proceed with semihard gluon
exchange and subsequent gluon bremsstrahlung a similar behaviour is
expected and the moments with cylindrical cut should approach a Poissonian
distribution. On the other hand, in collisions of heavy nuclei, if there is
a thermalization process with quark gluon plasma formation, the angular
ordering in the cascade process is destroyed and no Poissonian limit is
obtained. In this way the 
multiplicity moments could provide a new probe
of thermalization.

\end{document}